\documentstyle[preprint,aps,epsf]{revtex}

\newcommand{\referencestyle}{
\small
\abovedisplayskip=6pt
\belowdisplayskip=6pt
\vspace{12pt}}

\def\D{\Delta}

\def\Dt{\Delta\tau}

\def\t0{\tau_0}

\def\ben{\begin{eqnarray}}
\def\enn{\end{eqnarray}}
\def\ov{\over\displaystyle\strut}
\def\dst{\displaystyle\strut}
\def\l({\left(}
\def\r){\right)}

\begin{document}
\include{epsf}
\rightline{LUNFD6/(NFFL-7084) 1994}
\rightline{nucl-th/9408022}
\bigskip\bigskip\bigskip
\begin{center}
{ \Large\bf
Quantum Statistical Correlations for \\
Slowly Expanding Systems}
\end{center}
\bigskip
\begin{center}
	T. Cs\"org\H o$^{1,2}$\footnote{E-mails: csorgo@rmki.kfki.hu,
		csorgo@sunserv.kfki.hu \\
	\phantom{$^*$ E-mails:~} bengt@quark.lu.se \\
	\phantom{$^*$ E-mails:~} zimanyi@rmki.kfki.hu }
	B. L\"orstad$^{2,*}$ and J. Zim\'anyi$^{1,*}$
\end{center}
\medskip
\begin{center}
{\it
$^1$KFKI Research Institute for
Particle and Nuclear Physics of the \\
Hungarian Academy of Sciences,
H--1525 Budapest 114, P.O. Box 49. Hungary \\
$^{2}$Department of Elementary Particle Physics, Physics Institute, \\
University of Lund,
S\"olvegatan 14,
S - 223 62 Lund, Sweden
}
\end{center}
\vfill

\begin{abstract}
Quantum statistical correlations and momentum distributions are
calculated for a  spherically symmetric,
three-dimensionally expanding finite fireballs, for
non-relativistic expansions applying plane-wave approximation.
The new concepts of the geometrical temperature as well
as the thermal radius are presented  for the simplest
case.
The symmetry properties of the correlation function
are shown to reflect the symmetry properties of the
emission function.
\end{abstract}
\vfill
\rightline{Accepted for publication in Phys. Lett. B}

 \vfill\eject
Intensity interferometry has recently become a
popular tool to infer the freeze-out surface
heavy ion collisions and  to study the
space-time characteristics of the
strong interactions in elementary particle
reactions. The method was originally invented
{}~\cite{HBT} to measure the angular diameter of distant stars or galaxies,
the investigated objects being approximately static and the length-scales
astronomical. In principle the same method is applied
to measure the space-time characteristics of the
strong interaction, where the objects under study
are expanding systems, with life-times of a few fm/c ($10^{-23}$ sec)
and length-scales of a few fm ($10^{-15}$ m).

In recent publications~\cite{1d,3d} the new concept
 of the geometrical temperature and also the concept of
thermally generated length-scales~\cite{sinyukov,lutp,1d,3d}
were introduced for relativistically expanding systems both in
one and three dimensions. In ref.~\cite{3d}
application for recent
high energy heavy ion data from the CERN energy region
{}~\cite{NA35,NA44} was found: for three-dimensionally expanding,
cylindrically symmetric, finite systems the
Bose-Einstein Correlation Function (BECF)
was shown to measure the {\it thermal} length
scale, the region in the coordinate-space from where bosons
with similar momenta may emerge. The size of this region
is determined by the freeze-out temperature, the freeze-out time
and the flow gradients.  It was shown that this {\it thermal}
length-scale dominates the correlation function if the {\it
geometrical} characteristic length scales are sufficiently large.
However, the $m_t$ dependence of the parameters of the BECF
 were shown to contain correction terms
proportional to the ratio of the thermal/geometrical length scales.
As a complementary effect, the momentum distribution was shown
to be dominated by the {\it geometrical} and not the {\it thermal}
parameters, which provides us a possibility to measure the
(large) geometrical scales by a systematic study
of the single-particle momentum distribution.
These effects appear for  thermal lengths  small compared to
the geometrical size of the expanding system.
The standard results ~\cite{gyu_ka,zajc} are reproduced for the
case the geometrical length-scales being smaller than the thermal ones:
in this conventional case the BECF measures the geometrical size properly
and the freeze-out temperature can be extracted from the momentum
distribution.

The motivation for the present Letter is to investigate
how the thermal lengths and the geometrical temperature
reveal themself for the simplest systems where
they appear at all. Since the thermal length-scales appear due
the interplay of the flow-gradient and the temperature,
a non-relativistically expanding thermal source
with constant flow-gradient is the
simplest possible case. The key feature is that
the Bose-Einstein or Fermi-Dirac correlations appear
due to an enhanced/decreased probability of finding
identical particles in a similar momentum-state;
i. e. the correlations vanish for large momentum difference.
In case of an expanding source with flow gradients,
the mean momentum of the emitted particles changes
with changing locations. Locally, the width of the
momentum distribution of the emitted particles is determined
by the temperature; thus there will be a region in the
coordinate space, where the change in the mean momentum
of the emitted particles is of the same magnitude as the
width of the local momentum distribution. The size
of this region is referred to as the thermal length. If
it is smaller than the geometrical size of the
expanding system, the quantum statistical correlations
(Bose-Einstein or Fermi-Dirac) shall be sensitive
only to this thermal size, since this is the size
of the coordinate-space from which particles with
similar momenta may emerge.

Thus the simplest case is a non-relativistic
expansion with constant gradient.
We may choose the  case of the three-dimensionally expanding,
finite systems with spherical symmetry, which is
most probably realized in low energy heavy ion collisions.
For a recent review on the intensity interferometry
of low and intermediate energy heavy ion collisions
(E/A $<$ 1 GeV) see ref.~\cite{BGP}.

The particle emission is characterized by the emission function
$S(x,p)$ which is the probability that a particle is produced
at a given $ x = (t, \vec r\,) = (t,x,y,z)$ point in space-time
with the four-momentum $p = (E, \vec p\,) = (E, p_x, p_y, p_z)$,
where the particle is on mass shell, $ m^2 = E^2 - \vec p^{\,\, 2} $.
The quantum-mechanical analogy to the classical emission function
is the time-derivative of the Wigner-function,~\cite{pratt_csorgo}.

In terms of the time derivative of the
Wigner-function both the momentum spectra and
the quantum statistical correlation functions (QSCF-s) are prescribed.
A useful auxiliary function is the Fourier-transformed
emission function
\ben
\tilde S(\Delta k , K ) & = & \int d^4 x \,\,
		 S(x,K) \, \exp(i \Delta k \cdot x ),
\enn
where
\begin{eqnarray}
\Delta k  = p_1 - p_2, & \quad \mbox{\rm and}\quad &
K  = {\displaystyle\strut p_1 + p_2 \ov 2}
\end{eqnarray}
and $\Delta k \cdot x $ stands for the inner product
of the four-vectors.
Then the momentum distribution of the emitted particles,
$f(p)$ is given by
\ben
f(p) & = & \tilde S(\Delta k = 0, K = p),
\enn
which is normalized to unity,
\ben
\int d^3p \,\, f(p) & = & 1.
\enn
The two-particle Bose-Einstein or Fermi-Dirac
 correlation functions
 (BECF-s or FDCF-s) are prescribed in terms of our  auxiliary
function,
\ben
C(K,\Delta k) & = & 1 \pm {\displaystyle\strut
	 \mid \tilde S(\Delta k , K) \mid^2 \ov
			\mid \tilde S(0,K)\mid^2 },
\enn
where the $+$ sign stands for
bosons and the $-$ sign for fermions,
as was presented e.g. in ref. \cite{pratt_csorgo,zajc}.
In this Letter the effect of final state  Coulomb and Yukawa
interactions
shall be neglected as implicitly assumed by the above
equation, and the time-derivative of the Wigner-function
shall be approximated by classical emission functions.
We assume completely chaotic, thermal emission.
For Fermi-Dirac correlations, the plane-wave approximation is rather
crude especially due to the attractive final state
interactions in $pp$ and $nn$ systems for which systems
a large data set was taken at low and intermediate energies~\cite{BGP}.
The Coulomb repulsion in the $pp$ channel plays also an important role.
These final state interactions can be taken into account in the
Wigner-function formalism along the lines of ref.~\cite{pratt_csorgo}.
The Fourier-transformations which appear
in this Letter are the consequence of
the plane-wave approximation for the relative wave-function
$\psi_{\D k}(\vec r,t)$. They  correspond to integrals over
$\mid \psi_{\D k}(\vec r,t)\mid^2 = 1 + \cos(\D k \cdot x)$.
In the case the final state interactions are taken into account,
the Fourier-transformations are to be replaced by integral transformations
over the $\phi_{\D k}(\vec r, t)$ relative wave function
which can be calculated for the Coulomb + Yukawa final state interactions
as mentioned e.g. in ~\cite{pratt_csorgo}.

For central heavy ion collisions at low or intermediate
 energies the target and the projectile form a collective state
(stopping) which
can be described as a non-relativistically expanding fluid
within the framework of hydrodynamical models.
Due to the expansion, the fluid cools and at a certain temperature
it freezes out. In case of full stopping
and thermalization  (spherically symmetric momentum distribution)
the information about the initial direction is
lost thus the final freeze-out density distribution
becomes approximately
spherically symmetric, too.

We shall assume that the emission function is characterized by a given
a distribution  of production points $I(\vec r \,\,)$
and by a distribution of the freeze-out times, $H(t)$. The correlations between
space-time and momentum-space shall be introduced by a non-relativistic
momentum distribution. We assume that the expanding system
at the freeze-out is rare enough so that the quantum-statistical
single-particle distribution can be well approximated
by a Boltzmann-distribution,
\ben
	f_B(x;p) & = &
	 C \exp\l(- {\dst (\vec p - m \vec u(x))^2 \ov 2 m T}\r).
\enn
Here $\vec u(x) $ is the (non-relativistic) velocity,
the freeze-out temperature is denoted by $T$ and the
quantity $C = {\dst 1 \ov (2 \pi m T )^{3/2} }$ is a normalization constant.
Thus the emission function is characterized as
\ben
S(x;K) & = &  f_B(x;K) \,  I(\vec r\,) \,  H(t).
\enn
In order to simplify the results we keep only the mean value
and the width of the source distributions i.e. we shall
apply Gaussian approximations
for the distribution functions of $ t$ and $\vec r$ as follows
\ben
	I(\vec r\,) & = & {\displaystyle\strut 1\ov (2 \pi R_G^2 )}\,
		 \exp\l( - { \dst {\vec r\,}^2 \ov 2 R_G^2 } \r) \\
	H(t)  & = & {\dst 1\ov (2 \pi \D t^2 )^{1/2}} \,
			\exp\l( - {\dst (t - t_0)^2 \ov 2 \D t^2} \r).
\enn
We shall also briefly discuss our results for the general case.

We select a velocity of the 3D expanding matter at space-time point $x$
so that it be
spherically symmetric and  describe an expansion
in all three directions with a constant gradient. Thus the
velocity around the mean freeze-out time $t_0$  is assumed to
have the form
\ben
\vec u(x) & = & {\dst \vec r \ov t_0},
\enn
which describes a scaling solution of the non-relativistic
hydrodynamical equations at the mean freeze-out time $t_0$ for
$r << t_0$, ref.~\cite{jozso}.
Further we assume
that the freeze-out temperature in this non-relativistic case
is smaller than the mass of the particles, $ T < m $.

With the help of the above flow pattern, the Boltzmann
factor at the mean freeze-out time can be rewritten as
\ben
	J(\vec r - \vec r_a \, ) & = & f_B(t_0,\vec r \, ;K) \, = \,
	 C \exp\l( - {\dst \l(\vec r - \vec r_a(K) \r)^2 \ov 2 R_T^2  } \r)
\enn
i.e. the emission of the particles with momentum $K$ is
characterized by a Gaussian. The mean emission point $\vec r_a ( K)$ is
proportional to the observed momentum and the width  of the
emission points around the mean is determined by the
mean freeze-out time (inverse of the velocity gradient) and
the freeze-out temperature:
\ben
	\vec r_a (K)  =  {\dst t_0 \ov m} \vec K,
	& \qquad &      R_T^2  =  t_0^2 { T \ov m}.
\enn
This width $R_T$ shall be referred to as the 'thermal radius'.

When these approximations are valid,
our auxiliary function can be rewritten as
\ben
\tilde S(\D k, K) & = & \tilde H(\D E) \int d^3 \vec r \,
	 \exp\l(\, - i\, \D \vec k \, \vec r \,\r) \,\,
	 I(\vec r\,)\, \, J(\vec r - t_0 {\dst \vec K \ov m} ) ,\label{e:gen}
\enn
 where $\tilde H(\D E)$ stands for the Fourier-transformed
freeze-out time distribution.
Let's introduce the effective, $\vec K$ dependent
particle emission function
\ben
	I^{eff}_K (\vec r \,) & =& I(\vec r \,) \,\,
			 J(\vec r - t_0 {\dst \vec K \ov m} \, ).
\enn
Thus for arbitrary, Fourier-transformable distributions the
momentum distribution and the QSCF is given as
\ben
	{\dst d^3 n \ov dp_x dp_y dp_z} & \propto &
	f(\vec p\,) = \int d^3 \vec r \, \, I^{eff}_p (r)  \, = \\
	\null & = & \int d^3 \vec r \, \, I(\vec r \,) \,\,
		 J(\vec r - t_0 {\dst \vec p \ov m} \, ), \\
	C(\vec K, \Delta \vec k) & = & 1 \pm \mid \tilde H(\D E) \mid^2
		{\dst \mid \tilde I^{eff}_K(\D \vec k \,) \mid^2 \ov
		 \mid \tilde I^{eff}_K(\, 0 \,) \mid^2 }
\enn
Thus the momentum distribution is given by a convolution and the
QSCF shall be a symmetric function in the CMS of the pair,
where $\D E = 0$.

These general results take particularly simple form in the introduced
Gaussian approximations.
The auxiliary function $\tilde S(\D k,K)$ contains a factor which is the
 Fourier-transform of the product of Gaussians and so easily integrable.
The two Gaussian factors can be combined into a single Gaussian
by introducing the following notation:
\ben
	{\dst (\vec r - \vec r_a )^2 \ov R_T^2 } +
	 {\dst \vec r\,^2 \ov R_G^2 } & = & {\dst (\vec r - \vec r_b)^2 +
			\vec r_c\,^2 \ov R_*^2 } ,
\enn
and the new variables can be determined by the requirement that the
coefficients of the powers of $\vec r$ should be equal on both sides.
This yields the following relations:
\ben
	{\dst 1 \ov R_*^2 }& = &
	{\dst 1 \ov R_T^2 } + {\dst 1 \ov R_G^2 }, \label{e:r*} \\
	{\dst \vec r_a \ov R_T^2} & = & {\dst \vec r_b \ov R_*^2 }, \\
	{\dst \vec r_a\,^2 \ov R_T^2 } & = & {\dst \vec r_b\,^2 +
			 \vec r_c\,^2 \ov R_*^2 },
\enn
	and now the factor $\exp\l( - {\dst \vec r_c\,^2 \ov 2 R_*^2 }\r)$
	becomes independent of the integration. This factor shall enter
	the momentum distribution and can be rewritten as
\ben
	\exp\l( - {\dst \vec K^2 \ov 2 m T_*  }\r), & \quad\quad &
	\null
\enn
	where the effective temperature $T_*$ is the sum of the
	freeze-out temperature and the "geometrical temperature",
\ben
	T_*  =  T + T_G, & \qquad \mbox{where} \quad &
	T_G  =  m {\dst R_G^2 \ov t_0^2 }.\label{e:t*}
\enn
	The last line defines the geometrical temperature,
	being just the mass multiplied by the square of the mean expansion
	velocity ${\dst R_G \ov t_0}$.
	The geometrical temperature $T_G$ is related to the geometrical
	radius $R_G$ exactly the same way as the freeze-out temperature
	$T$ is related to the thermal radius $R_T$:
\ben
	R_G^2  & = & t_0^2 {\dst T_G \ov m}, \\
	R_T^2  & = & t_0^2 {\dst T \ov m}.
\enn

	The remaining integral in the auxiliary function
	is trivial, resulting in the following momentum-distribution
	and correlation function:
\ben
	f(\vec p\,) & = &  {\dst 1 \ov (2 \pi m T_* )^{3/2} }
	\exp\l( - {\dst \vec p\,^2 \ov 2 m T_*} \r), \\
	C(\vec K, \D \vec k \, ) & = &
		1 \pm \exp( - R_*^2 \D \vec k^2) \,
		    \exp( - \D t^2 \D E^2) .
\enn
	The effective  temperature $T_*$ and the effective radius parameter,
	$R_*$ were defined in eqs.~\ref{e:r*},\ref{e:t*}.

	The relative momentum $\D \vec k$,
	which enters the  correlation function, is invariant
	under Galilei-transformations. However,
	the energy difference is
	not invariant even under the non-relativistic Galilei transformations.
	Let us re-formulate $\D E$ in such a way that
	its  specific directional dependence be more transparent.
	The energy difference could be identically rewritten as
\ben
	\D E^2  =  {\dst \vec p_1\,^2 - \vec p_2\,^2 \ov 2 m } & = &
	\l( {\dst \vec p_1 + \vec p_2 \ov 2 } \r) { \dst \D \vec k \ov m} =
	\vec v_{K} \D \vec k,
\enn
	where we have introduced the mean velocity of the
	pair,
	$\vec v_K ={\dst \vec p_1 + \vec p_2 \ov 2 m} = {\dst \vec K \ov m}$.

	Let us define the {\it out} direction to be parallel to the mean
	velocity of the pair, $\vec v_K$, and the {\it perp} index shall
	be used to index the remaining two principal directions,
	both being perpendicular to the {\it out} direction.
	This naming convention corresponds to the one used in high
	energy heavy ion collisions~\cite{bertsch,lutp}.

	By this definition we evaluate the correlation function
	in  a frame which is assigned to a given mean momentum of the
	particle pair and {\it not} that of the fireball:
	by changing the mean momentum
	 we  change our reference frame too.
	In this co-moving frame the relative momentum is decomposed as
\ben
	\D \vec k = (Q_{out},\vec Q_{perp} )
\enn
	and the correlation function can be rewritten as
\ben
	C(Q_{out},Q_{perp}) & = & 1 \pm
	 \exp\l( - R_{perp}^2 Q_{perp}^2 - R_{out}^2 Q_{out}^2 \r), \\
	R_{perp}^2 & = & R_*^2 = {\dst R_T^2 R_G^2 \ov R_T^2 + R_G^2 }, \\
	R_{out}^2 & = & R_*^2 + \vec v_K\,^2 \D t^2. \label{e:out}
\enn
	Thus the perpendicular components shall measure the effective
	radius $R_*$. What is the interpretation of this quantity?

	There are two length-scales in the problem: the geometrical
	length-scale $R_G$ and the thermal one $R_T$, the latter being
	generated by the flow-gradient and the temperature.
	The effective $R_*$
	is dominated by the smaller
	of the two, cf. eq.\ref{e:r*}.
	 With other words, for large, relatively cold
	three-dimensionally expanding systems not the whole source can be
	seen by quantum statistical correlations, but only a part of the
	whole system, which is characterized by a thermal length-scale $R_T$.
	This corresponds to the limiting case $ R_T << R_G$
	(i.e. $T << T_G$), resulting in a thermally dominated
	correlation function and a geometrically dominated momentum
	distribution:
\ben
	R_{perp}^2 & = & R_T^2, \\
	R_{out}^2 & = & R_T^2 + \vec v_K \,^2 \D \tau^2, \\
	T_* & = & T_G.
\enn
	The region, viewed by the QSCF is shown also on Fig.1.
	The generation of the effective, geometrical
	temperature for cold, expanding systems is illustrated on Fig. 2.

	It is interesting to investigate the other limiting case when
	$R_T >> R_G $.
	In this case we re-obtain the standard results:
\ben
	R_{perp}^2  & = &  R_G^2, \\
	R_{out}^2 &  = &  R_G^2 + \vec v_K\,^2 \Dt^2,\\
	T_* & = & T,
\enn
	i.e. if the thermal length scale is larger than the geometrical
	size in all three direction, the BECF measurement determines
	the geometrical sizes properly, and the momentum
	distribution will be determined by the freeze-out temperature
	giving a thermal distribution for a static source.
	The region seen by the QSCF is shown on Fig. 3.

	From eq. \ref{e:out} it follows that the out component
	in general shall be sensitive to the duration of the
	freeze-out time distribution, since it contains a term
	$\vec v_K\,^2 \D t^2 $. This term vanishes in the
	center of mass system  (CMS) of the particle pair,
	since $\D E = \vec v_K = 0$ in CMS.
	The correlation function in the considered case becomes symmetric,
	when evaluated in the CMS of the pair, since
\ben
	Q_I^2 = Q_{perp}^2 + Q_{out}^2
		\qquad \mbox{\rm in~CMS~of~the~pair}
\enn
	is the invariant momentum difference.
	The correlation function shall take up the following,
	particularly simple form:
\ben
	C(Q_I) & = & 1 \pm \exp( - R_*^2 Q_I^2)
		\qquad \mbox{\rm in~CMS~of~the~pair}.
\enn
	This is to be contrasted to the three dimensionally expanding,
	relativistic, cylindrically symmetric case where the correlation
	function was found to be symmetric in the longitudinally
	comoving system (LCMS) of the particle pair~\cite{3d}, and {\it not}
	in their CMS.

	This result emphasizes the importance of the analysis of
	the symmetry properties of the correlation functions (CF-s),
	since in two cases we found that these properties reflect the
	symmetry properties of the emission pattern: an axially symmetric
	3d emission generates a CF symmetric in LCMS while a spherically
	symmetric 3d emission function
	generates a CF symmetric in the CMS of the pair.
	In case the emission pattern describes a 1d expansion~\cite{1d,lutp},
	the symmetry of the BECF is broken. Note that the symmetry
	appears in  the 3d cases when the CF-s are dominated by the
	{\it thermal} length-scales and obviously a thermal distribution
	becomes symmetric if one can find the appropriately
	comoving reference frame.

	We have three input parameters:
	$R_G, t_0 $ and $T$ which determine two measurables,
	$R_*$ and $T_*$. Thus the geometrical parameters can be determined
	only as a function of one (within the model unknown) parameter,
	which we may choose to be $T$.

	In any case, bounds can be given since
\ben
	t_0^2 & = & R_*^2 {\dst T_* m \ov T (T_* - T) }
			\,\, \ge \,\, 4 \,\, R_*^2 \,  {\dst m \ov T_*}
			\label{e:rlim} \\
	R_G^2 & = & R_*^2 \, {\dst T_* \ov T} \,\, \ge \,\, R_*^2 \label{e:tlim}
\enn
	Our model approximations are valid if
	$ m   >  T_* $
	which correspond to the requirement $t_0 > R_*$.

 {\it In summary}, we have calculated the momentum distribution
and the quantum statistical (Bose-Einstein or Fermi-Dirac)
 correlation function for
three-dimensionally expanding systems
in a non-relativistic model for spherically symmetric expansions,
applying plane-wave approximation for a completely chaotic (thermal)
source.

 In all the three principal
directions, two length-scales are present:
the geometrical and the thermal one.
The latter is due to the combined effect
 of the flow-gradient and the temperature within the considered
model for particle emission.
	The radius parameters of the correlation function are
	dominated by the {\it shorter} of the thermal
	and geometrical length scales.
	A complementary effect is that
	the effective temperature is a sum of the freeze-out temperature
        and the
	geometrical temperature. Thus the effective temperature
        is dominated
	by the {\it higher} of the two temperature scales.

In the chosen case the Bose-Einstein (Fermi-Dirac) correlation functions
were found to be symmetric in the center of mass system of the pair,
had the geometrical sizes been
larger than the thermal length-scale. Lower limits on the
geometrical size and the freeze-out time
can be obtained from a simultaneous analysis
of the {\it momentum distribution} and the correlation function, eqs.
\ref{e:tlim}, \ref{e:rlim}.

We have given examples to show that an analysis of the symmetry properties
of the quantum statistical correlation function
might indicate the symmetry properties of the
underlying flow pattern and density distribution.

{\it Acknowledgments:}
Cs.T. would like to thank G. Gustafson for helpful discussions,
and to M. Gyulassy for stimulating discussions and for
kind hospitality at Columbia University in August 1993.
This work was supported
by the Human Capital and Mobility (COST) programme of the
EEC under grant No. CIPA - CT - 92 - 0418
(DG 12 HSMU), by the Hungarian
NSF  under Grants  No. OTKA-F4019 and OTKA-T2973,
and is supported by the USA-Hungarian Joint  Fund.

\vfill
\eject
\begin{center}
Figure Captions
\end{center}
\begin{itemize}
\item{Figure 1.} Illustration of the region seen by the
	correlation function in case $R_T << R_G$. The
	correlation function views the shaded area.
	In the general case the correlation function yields a radius $R_*$,
	given by eq.~\ref{e:r*}.

\item{Figure 2.} The generation of the effective temperature
	in case $R_T << R_G$. The particles at position
	$\vec r$ are emitted with a mean momentum $\vec K = m \vec r / t_0$.
	Since $T << T_G$ in this case, the density distribution
	in space is mapped to a density distribution in momentum
	space. Thus the effective temperature shall be given by
	$T_G = m R_G^2 / t_0^2$.

\item{Figure 3.} Illustration of the region seen by the
	correlation function in case $R_T >> R_G$. The
	correlation function views the shaded area.
	In the general case the correlation function yields a radius $R_*$,
	given by eq.~\ref{e:r*}.

\end{itemize}
\vfill\eject
\vfill
\begin{figure}
          \begin{center}
          \leavevmode\epsfysize=4.5in
          \epsfbox{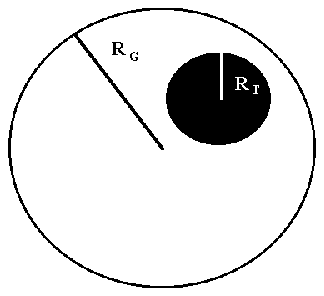}
          \end{center}
\caption{
}
\end{figure}
\vfill\eject
\begin{figure}
          \begin{center}
          \leavevmode\epsfysize=4.5in
          \epsfbox{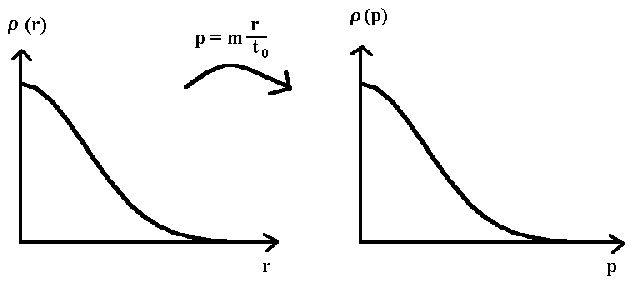}
          \end{center}
\caption{
}
\end{figure}
\vfill\eject
\begin{figure}
          \begin{center}
          \leavevmode\epsfysize=4.5in
          \epsfbox{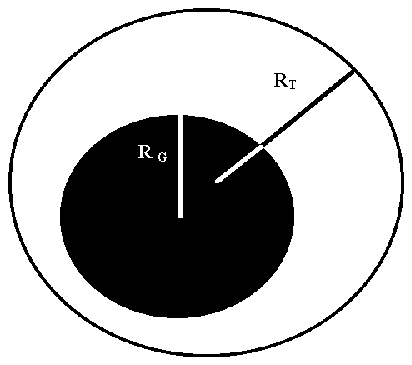}
          \end{center}
\caption{
}
\end{figure}
\vfill\eject

\end{document}